\documentclass[aps,pre,twocolumn,groupedaddress,superscriptaddress,showpacs]{revtex4-2}
\usepackage{epsfig,amssymb,amsmath,graphicx,subfigure,hyperref}
\usepackage{tabularx}
\usepackage{float}
\usepackage{setspace}
\usepackage{color}
\usepackage{array} 
\newcommand{\PreserveBackslash}[1]{\let\temp=\\#1\let\\=\temp} \newcolumntype{C}[1]{>{\PreserveBackslash\centering}p{#1}} \newcolumntype{R}[1]{>{\PreserveBackslash\raggedleft}p{#1}} \newcolumntype{L}[1]{>{\PreserveBackslash\raggedright}p{#1}} 
\usepackage{graphicx}
\usepackage{dcolumn}
\usepackage{bm}

\begin{document}
\preprint{APS/123-QED}
\title{Effects of orientational and positional randomness of particles on photonic band gap}
\author{Zichen Qin}
\affiliation{School of Physics and Technology, Wuhan University, Wuhan 430072, China}
\author{Tao Liu}
\affiliation{School of Physics and Technology, Wuhan University, Wuhan 430072, China}
\author{Duanduan Wan}
\email[E-mail: ]{ddwan@whu.edu.cn}
\affiliation{School of Physics and Technology, Wuhan University, Wuhan 430072, China}

\begin{abstract}
A recent work [PRL, 126, 208002 (2021)] has explored how thermal noise-induced randomness in a self-assembled photonic crystal affects photonic band gaps (PBGs). For the system of a two-dimensional photonic crystal composed of a self-assembled array of rods with square cross sections, it was found that its PBGs can exist over an extensive range of packing densities. Counterintuitively, at intermediate packing densities, the transverse magnetic (TM) band gap of the self-assembled system can be larger than that of its corresponding perfect system (rods arranged in a perfect square lattice and having identical orientations). Due to shape anisotropicity, the randomness in the self-assembled system contains two kinds of randomness, i.e., positional and orientational randomness of the particles. In this work, we further investigate how PBGs are influenced solely by positional or orientational randomness. We find that compared to the perfect situation, the introduction of only orientational randomness decreases the transverse electric (TE) band gap while having no obvious effects on the transverse magnetic (TM) band gap. In contrast, the introduction of only positional randomness decreases the TE band gap significantly, while it can widen or narrow the TM band gap, depending on the parameter range. We also discuss the thermal (i.e.,~self-assembled) system where two kinds of randomness are present. Our study contributes to a better understanding of the role orientational randomness and positional randomness play on PBGs, and may benefit the PBG engineering of photonic crystals through self-assembly approaches.
\end{abstract}

\maketitle

\section{Introduction}
The self-assembly feature of colloids, i.e., colloidal particles arrange themselves into an ordered structure, has intrinsic physical meaning and potential application in engineering functional materials \cite{Manoharan2015, Boles2016}. Recent advances in synthesis have allowed the realization of various anisotropic particles \cite{Henzie2012, Young2013, Gong2017, Forster2011, Hosein2010, Meijer2017, Miszta2011, Kraft2012, He2020}, which leads to a diverse range of self-assembled superlattices. Simulations of hard colloids (e.g., Refs.~\cite{Gang2011, Haji-Akbari2009, Torquato2009, Agarwal2011, Damasceno2012_science, Ni2012, Marechal2010, Chen2014, Wan2018, Wan2022, Klotsa2018, Wan2019}) predict complex structures from an even larger variety of anisotropic shapes. When interparticle distances commensurate with wavelengths of light,  colloidal self-assembly has been explored as a route to fabricate photonic crystals with photonic band gaps (PBGs)  (e.g., Refs.~\cite{Ye2001,Vlasov2001, Hosein2010, Forster2011,Sowade2016}). This bottom-up method has advantages such as low cost, low energy consumption, adjustable PBG frequency through particle size, and crystals can be produced over large areas \cite{Moon2010, Kim2011, Zhao2014}. Photonic crystals from self-assembly approaches have been realized experimentally and proposed theoretically (e.g., Refs.~\cite{Ding2014, Hynninen2007, Woldering2011, Pattabhiraman2017, Wang2017, Changizrezaei2017, Lei2018}).

An inevitable question about self-assembly approaches is how ``randomness" arising from thermal noise in a self-assembled colloidal structure affects its PBGs. A recent study explores two-dimensional self-assembled structures, i.e., parallel dielectric rods of infinitely long length with circular or square cross section, and finds that in general, the randomness in the self-assembled (i.e.,~thermal) structure decreases the PBGs, compared to the system of rods arranged in its corresponding perfect lattice \cite{Wan2021}. However, for rods with square cross sections at intermediate packing densities, interestingly, the transverse magnetic (TM) band gap of the self-assembled system can be larger than that of rods arranged in a perfect square lattice. As this counterintuitive observation only exists for rods with square cross sections but not for rods with circular cross sections, we expect it can be related to the shape anisotropicity of the square cross sections. As shape anisotropicity leads to the existence of two kinds of randomness (i.e.,~orientational randomness and positional randomness) of the particles in the self-assembled system, it remains unclear how orientational or positional randomness alone will affect PBGs. 

\begin{figure}
\includegraphics[width=0.47\textwidth]{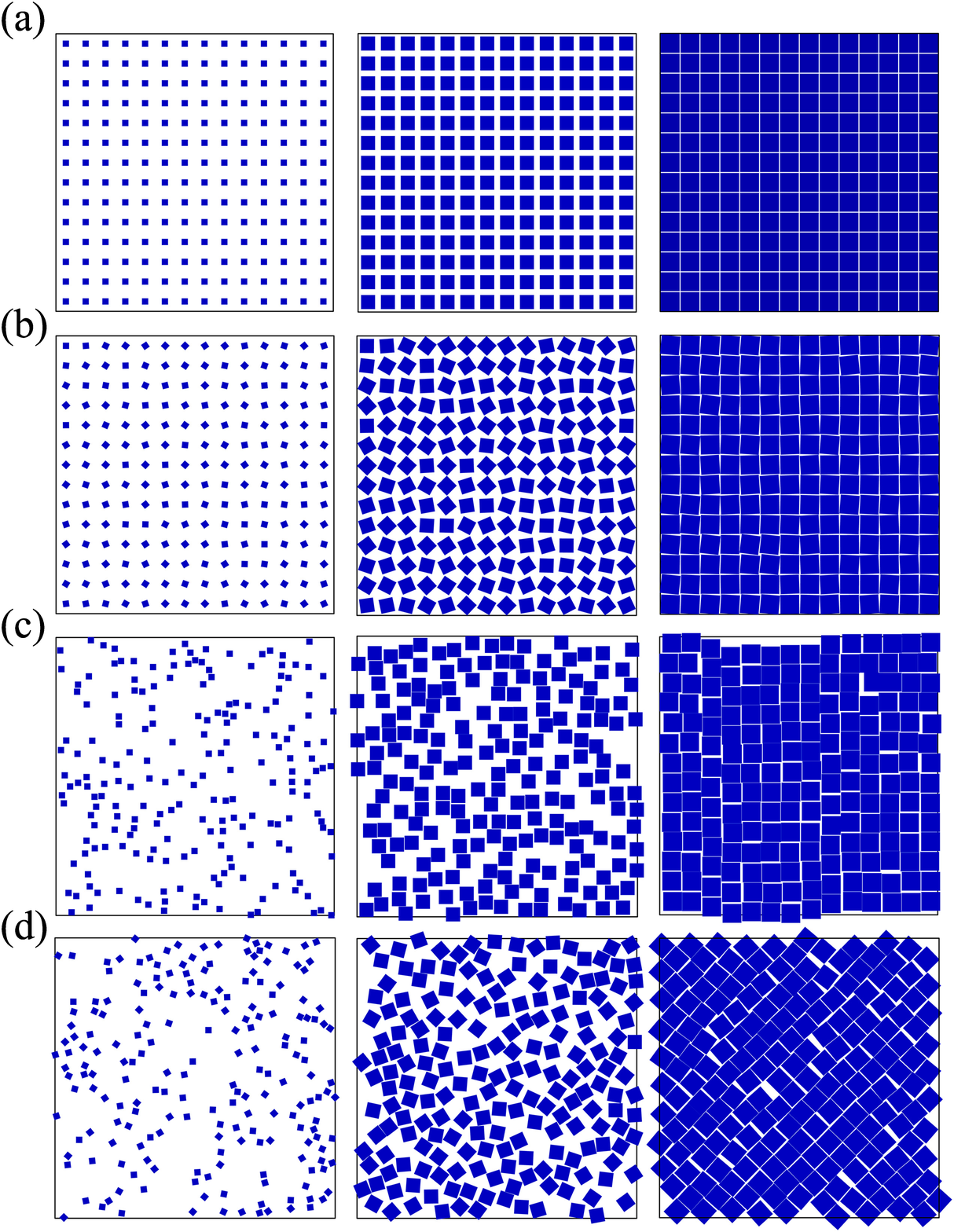}
\caption{
Snapshots of equilibrium configurations of $N=196$ rods with square cross sections in the four situations. Blue squares represent the cross sections of the dielectric rods, while white is air. (a) The perfect situation. (b) The random orientation situation. (c) The random position situation. (d) The thermal situation. Packing densities are $\phi =0.1, 0.5, 0.9$ (from left to right).
}
\label{illustration}
\end{figure}

To pursue an answer to this question, in this work, we do a further investigation on the rods with square cross sections system. In particular, we consider four situations: the perfect situation where rods are arranged in a perfect square lattice and have no rotations, the random orientation situation where rods are in the perfect square lattice as well but are random in orientations, the random position situation where rods are random in position while having identical orientations, and the thermal situation where rods random in both orientations and positions (see examples in Fig.~\ref{illustration}). We find that compared to the perfect situation, the introduction of only orientational randomness decreases the transverse electric (TE) band gap while having no apparent effects on the transverse magnetic (TM) band gap. In contrast, the introduction of only positional randomness decreases the TE band gap significantly, while it can widen or narrow the TM band gap, depending on the parameter range. We also discuss the thermal (i.e.,~self-assembled) system where both two kinds of randomness are present. This work contributes to a better understanding of the role orientational and positional randomness play on PBGs, and may further benefit the PBG engineering of photonic crystals through self-assembly approaches. 

\begin{figure}
\includegraphics[width=0.45\textwidth]{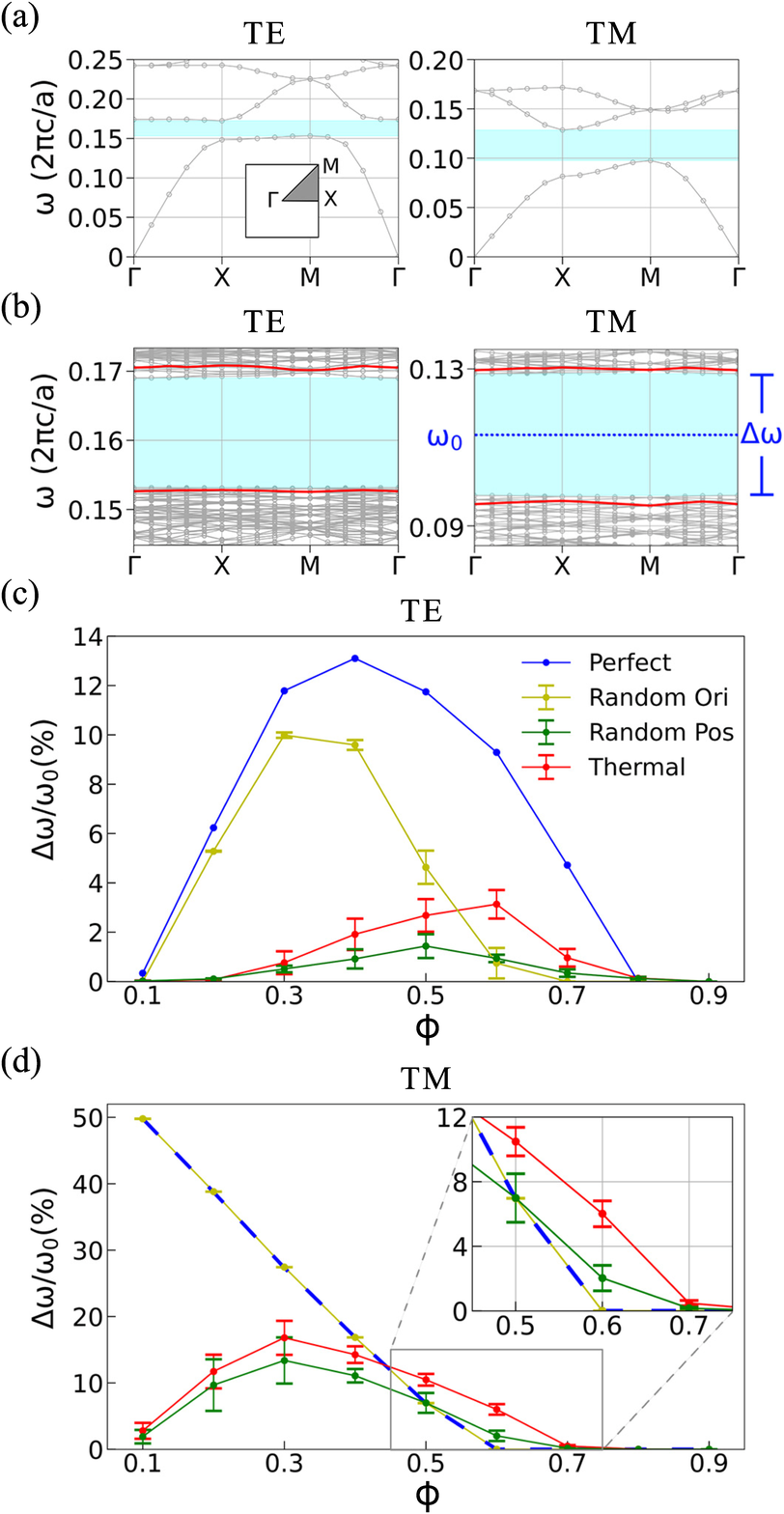}
\caption{
Photonic band structure and band gap for arrays of parallel dielectric rods of square cross sections with side length $a$ ($a$ is also the length unit) embedded in air ($\epsilon=1$). (a) An example of the band structure of the TE and TM modes in the perfect situation, with $\epsilon=20$ and $\phi=0.3$. The inset shows the first Brillouin zone for the periodic structure, with the symmetry points indicated. The light blue area represents the band gap observed. (b) Parallel results for snapshots of $N=196$ rods in the random orientation situation, with parameters the same as those in (a). The red lines indicate the $(N-5)$th and the $(N+6)$th bands. The band gap (light blue area) reported is the largest gap among all the gaps between these two bands. The central frequency of the gap is denoted as ${\omega}_0$, and the width of the gap is $\Delta \omega$ (blue labels). (c-d) Gap size $\Delta \omega / {\omega}_0$ as a function of  $\phi$ for the TE mode (c) and TM mode (d), with $\epsilon=20$. The inset in (d) is a zoomed plot. Except for the perfect situation, the gap size is obtained by averaging over five independent simulation snapshots. Error bars indicate the standard deviation.
}
\label{comparison}
\end{figure}

\section{Methods}

We perform Monte Carlo (MC) simulations with periodic boundary conditions to generate the two-dimensional structure of $N=196$ rods with square cross section, using the hard particle Monte Carlo module in {\scriptsize HOOMD-blue} \cite{hoomd, Glaser2015}. For the random orientation situation, at a given packing density we placed the centers of the rods in a perfect square lattice, and then randomly rotated the rods using MC steps. For the random position situation and the thermal situation, we initialized the system at a low packing density, slowly compressed it to a target packing density using MC steps, and then equilibrated it. The difference is that for the random position situation, only translational MC moves were needed. See Refs.~\cite{Haji-Akbari2009, Wan2021} for more simulation details.  We used the supercell method implemented in the open source code {\scriptsize MIT PHOTONIC BANDS} \cite{Johnson2001} to obtain the photonic band structure of equilibrated snapshots. We also used the finite element package COMSOL Multiphysics to calculate the band structure of a few samples for double check. In practice, to take into consideration of the existence of possible defect modes, instead of the band gap between the $N$th and $(N+1)$th bands, we report the band gap $\Delta \omega$ which has the largest size among all gaps between the $(N-5)$th and the $(N+6)$th bands, where $N$ is the particle number.

\section{Results and discussion}

Figure \ref{illustration} shows equilibrium configurations of $N=196$ rods with square cross sections at a few selected packing densities in the four abovementioned situations. In the perfect situation, rods have no positional or orientational randomness (Fig.~\ref{illustration}(a)). In the random orientation/position situation, rods have only orientational/positional randomness (Fig.~\ref{illustration}(b) and (c)). In the thermal situation, rods have two kinds of randomness (Fig.~\ref{illustration}(d)).

\begin{figure}
\includegraphics[width=0.42\textwidth]{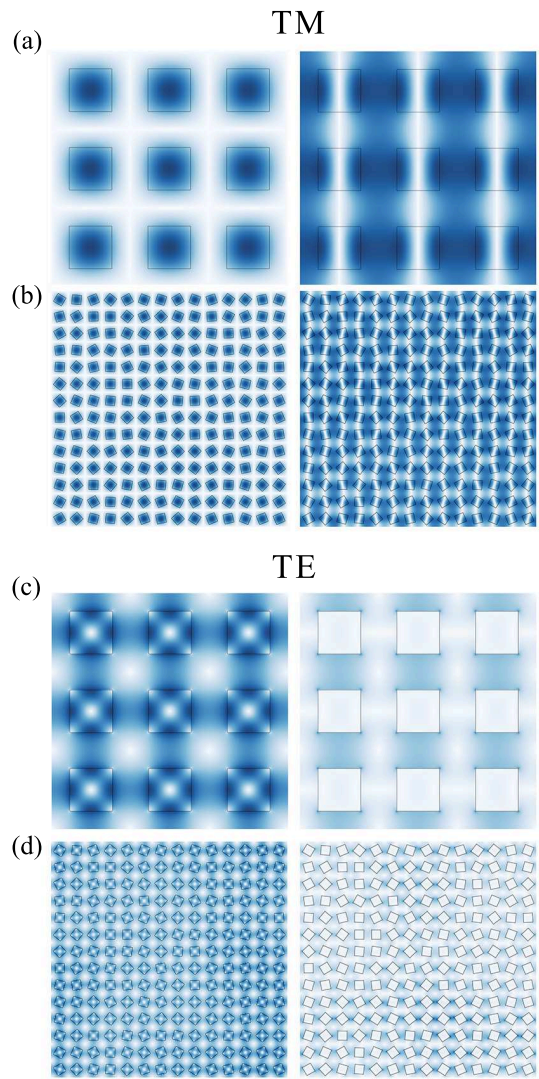}
\caption{
Electric field magnitude for TM (a and b) and TE (c and d) polarization, with $\epsilon=20$ and $\phi=0.3$. (a) At the M-point of the first band (left), and the X-point of the second band (right) in the perfect situation (Fig.~\ref{illustration}(a), right). It shows $3\times 3$ periods for a better view. (b) Examples of modes before (left) and after (right) the gap in the random orientation situation with $N=196$ (Fig.~\ref{illustration}(b), right). (c-d) Same as (a) and (b), for TE polarization.
}
\label{electric_field}
\end{figure}

\begin{figure*}
\includegraphics[width=0.86\textwidth]{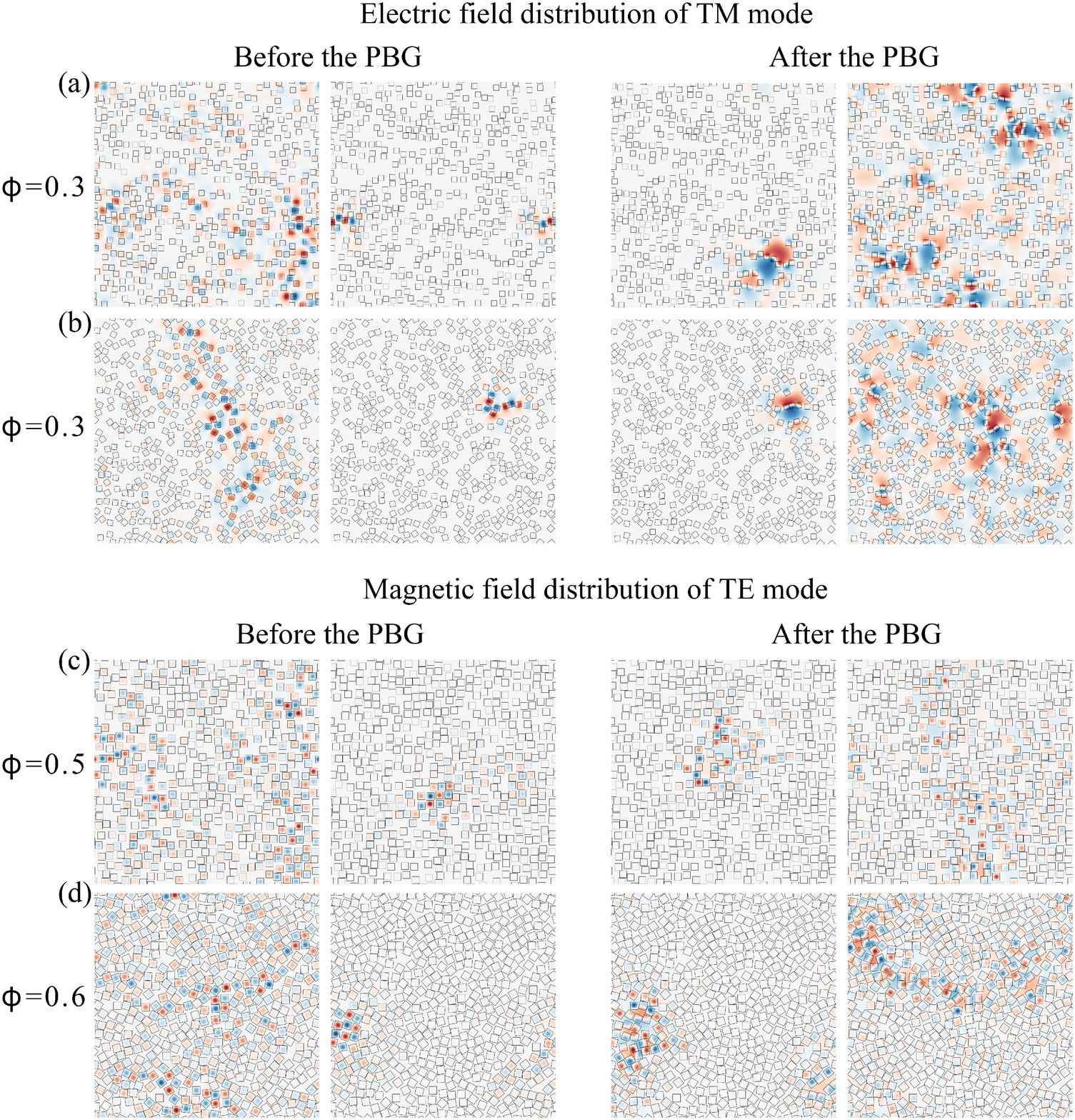}
\caption{
Electromagnetic field distribution of the random position and the thermal situation in a system of $N=500$ rods. The figures from left to right are extended mode before the PBG, localized mode before the PBG, localized mode after the PBG, and extended mode after the PBG, respectively. (a) The random position situation with $\phi=0.3$. (b) The thermal situation with $\phi=0.3$. (c) The random position situation with $\phi=0.5$. (d) The thermal situation with $\phi=0.6$.
}
\label{field_distribution}
\end{figure*}

\begin{figure*}
\includegraphics[width=\textwidth]{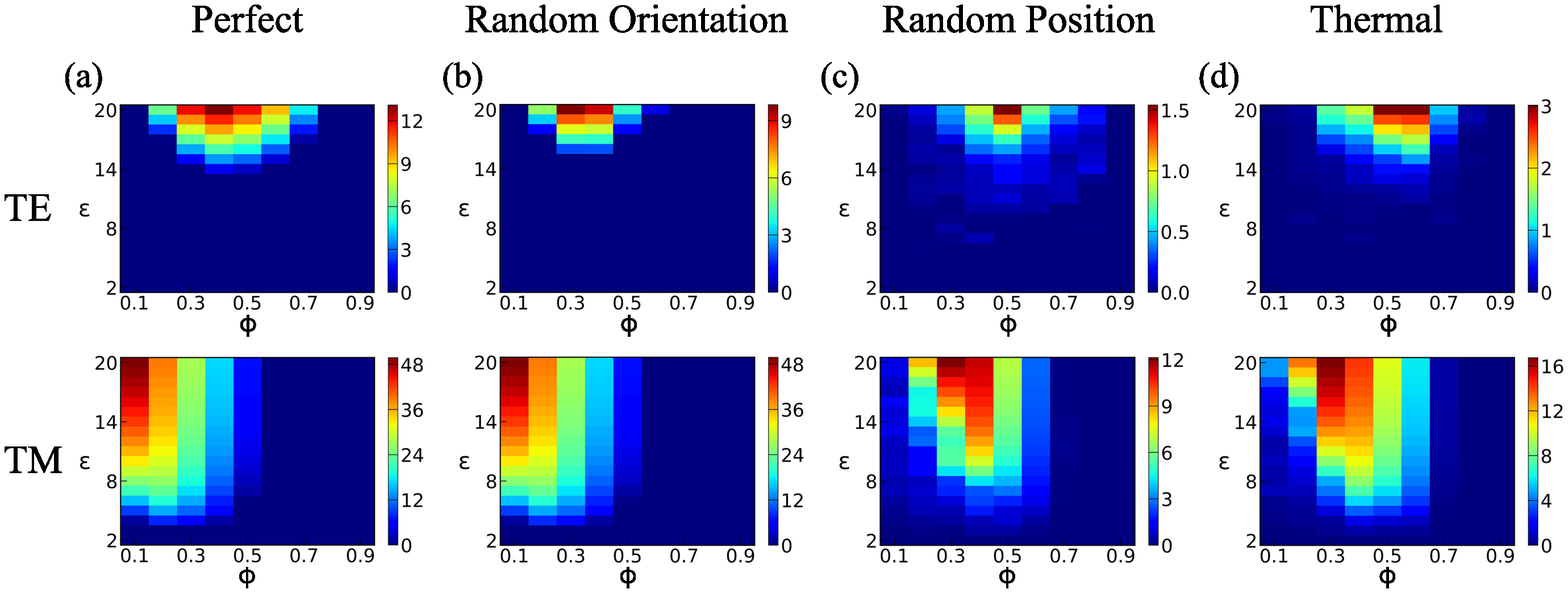}
\caption{
Band gap size (the percentage of $\Delta \omega / {\omega}_0$) as a function of $\phi$ and $\epsilon$ for the TE and TM modes. (a) The perfect situation. (b) The random orientation situation. (c) The random position situation. (d) The thermal situation.
}
\label{gap_size}
\end{figure*}

In order to study the influence of positional and orientational randomness on PBGs, we first plot the packing density dependence of relative gap size (defined as $\triangle \omega/\omega_{0}$, with $\triangle\omega$ the width of the PBG and $\omega_{0}$ the central frequency) at dielectric constant $\epsilon =20$ in the four situations (see Fig.~\ref{comparison}). We choose $\epsilon =20$ to be consistent with the previous study \cite{Wan2021}. We see that the four lines are basically divided into two groups: The upper two lines represent the perfect situation and the random orientation situation, while the lower two lines represent the random position situation and the thermal situation. For TE polarization, the perfect situation always has the largest PBG. On the basis of the perfect situation, the introduction of orientational randomness, which gives the random orientation situation, is represented by the line just below the perfect situation's. As we turn to see the lower two lines, we find that the random position situation has a smaller PBG, and when we again added orientational randomness and got the thermal situation, it gives a larger PBG, represented by the line just above the random position situation's. It is interesting to see that when the rods are placed in a perfect lattice, the introduction of orientational randomness may cause a loss of PBG. However, when it comes to a system with positional randomness, the introduction of orientational randomness can always give rise to an increase in PBG. 

For TM polarization, the situation is similar to the TE polarization, with another intriguing fact that the perfect situation and the random orientation situation have almost the same PBG at all packing densities, which is shown as two overlapped lines in Fig.~\ref{comparison}(d). Besides, although at most packing densities the perfect situation has the largest band gap size, at packing densities near $\phi=0.6$, it can be smaller than the thermal situation and even the random position situation, consistent with previous finding \cite{Wan2021}. 

To gain insight into why the perfect and random position situations have almost the same gap size for TM polarization while not for TE polarization, we plot the electric field magnitude of modes near the PBG in these two situations in Fig.~\ref{electric_field}. For TM polarization, we observe in the perfect situation, before the PBG, the electric field is more concentrated in a circle inside the square cross section of the particle, and decreases gradually as it goes away from the center (Fig.~\ref{electric_field}(a), left). The electric field distribution around a particle is similar for modes before the band gap in the random orientation situation, and the rotation of particles has little effect on the overall field distribution as the electric field is continuous across the rod boundary (Fig.~\ref{electric_field}(b), left). For modes after the PBG, the field distribution in these two situations also shows similar characters, i.e., the field distribution on every particle is dipole-like and the overall distribution shows stripes of strong and weak areas (Fig.~\ref{electric_field}(a) and (b), right). In contrast, for TE polarization, the electric field is not continuous in the direction perpendicular to the rod boundary, and thus the orientation of particles has more significant influence on the overall field distribution, e.g., when two particles are more vertex-to-vertex aligned, the field in between is enhanced (Fig.~\ref{electric_field}(d), right). 

For the random position and the thermal situations (the lower two lines in Fig.~\ref{comparison} (c) and (d)), to confirm the existence of the band gap, we show the electric field distribution of some TM modes around the PBGs in Fig.~\ref{field_distribution}(a) and (b), and the magnetic field distribution of some TE modes around the PBGs in Fig.~\ref{field_distribution}(c) and (d), respectively. The existence of localized and extended modes around a PBG helps to demonstrate the PBG. Localized and extended modes around a PBG have also been observed in other 2D systems (e.g., Refs.~\cite{Florescu2009, Ricouvier2019}). 

To better understand the random orientation and the random position situations, we investigate a wider range of $\epsilon$. Figure \ref{gap_size} plots the relative gap size as a function of $\phi$ and $\epsilon$ in the four situations, with $\phi \in [0.1, 0.9]$ as that in Fig.~\ref{comparison}, and $\epsilon \in [2, 20]$. As can be seen from the figure, at a same $\phi$ value, the gap size increases as $\epsilon$ increases, and $\epsilon=20$ gives the largest gap size. A detailed comparison of the gap sizes shows that the results are similar to that for $\epsilon=20$. For TE polarization, the introduction of positional randomness and orientational randomness always causes a reduction of PBG, and when we together introduce these two kinds of randomness, its gap size will always be larger than the one of positional randomness, and at some packing densities may be beyond the one of orientational randomness, however will still be below the gap size of the perfect situation. For TM polarization, the random orientation and the perfect situation have almost the same band gap size at all studied parameter values. The random position situation at most packing densities has a smaller PBG than the perfect one, but at some intermediate packing densities can be larger than it. For the thermal situation, just like what we find in TE polarization, its PBG is always larger than the random position situation.

\section{Conclusions}
 
In summary, we studied photonic band gaps in disordered 2D photonic crystals of dielectric rods with square cross sections with two kinds of typical randomness of particle position and orientation. We find that the band gap size is far more sensitive to disorder with positional randomness than with orientational randomness. On this basis, we may consider the orientational randomness as a perturbation. However, this perturbation can lead to different effects, which may even be quite the opposite, under diverse conditions. For TE polarization of the perfect situation, this perturbation can give rise to a reduction of the PBG, while for both TE and TM polarization of the random position situation, it always results in an expansion. And for TM polarization of the perfect situation, this perturbation has no apparent impact on the gap size. As positional and orientational randomness is present for systems with any anisotropic particle shape, studies on other particle shapes can be carried out in the future. We hope our work provides insight into understanding the PBG dependence on positional and orientational randomness, and may further benefit the PBG engineering of photonic crystals through self-assembly approaches.

\acknowledgments  
This work was supported by the National Natural Science Foundation of China (Grant No.~12274330, 11904265), and Knowledge Innovation Program of Wuhan-Shuguang (Grant No.~2022010801020125).

\bibliography{pc_refs}

\end{document}